\documentclass{ifacconf}

\usepackage{graphicx}      % include this line if your document contains figures
\usepackage{natbib}        % required for bibliography
\usepackage{amsmath}

\begin{document}
\begin{frontmatter}

\title{Variational Mode Decomposition as Trusted Data Augmentation in ML-based Power System Stability Assessment\thanksref{footnoteinfo}} 
% Title, preferably not more than 10 words.

\thanks[footnoteinfo]{This research was funded in part by NYSERDA under agreement 137951 and the National Science Foundation, Grant No. 2231677.}

\author[First]{Tetiana~Bogodorova} 
\author[Second]{Denis~Osipov} 
\author[First]{Luigi~Vanfretti}

\address[First]{Electrical, Computer, and Systems Engineering Department, Rensselaer Polytechnic Institute, Troy, NY 12180 USA, (e-mail: bogodt2@rpi.edu; vanfrl@rpi.edu).}
\address[Second]{System Planning \& Analysis, New York Power Authority, 
   Albany, NY 12207, USA (e-mail: Denis.Osipov@nypa.gov)}
%\address[Third]{Electrical Engineering Department, 
%   Seoul National University, Seoul, Korea, (e-mail: author@snu.ac.kr)}

\begin{abstract}                % Abstract of 50--100 words
Balanced data is required for deep neural networks (DNNs) when learning to perform power system stability assessment. However, power system measurement data contains relatively few events from where power system dynamics can be learnt. To mitigate this imbalance, we propose a novel data augmentation strategy preserving the dynamic characteristics to be learnt. The augmentation is performed using Variational Mode Decomposition. The detrended and the augmented data are tested for distributions similarity using Kernel Maximum Mean Discrepancy test. In addition, the effectiveness of the augmentation methodology is validated via training an Encoder DNN utilizing original data, testing using the augmented data, and evaluating the Encoder's performance employing several metrics.
\end{abstract}

\begin{keyword}
Convolutional neural networks, data augmentation, deep learning, power system stability assessment, variational mode decomposition.
\end{keyword}

\end{frontmatter}
%===============================================================================

\section{Introduction}

Power grid measurements, even if regularly collected and stored, do not contain enough event recordings from where dynamics can be learnt. This poses a challenge to train a deep neural networks (DNNs) to perform power system security assessment tasks that require to recognize, predict or classify power system dynamics with high accuracy ~\cite{narasimham_arava_analyzing_2018}. Thus, the hybrid solution of using synthetic and real measurements mixture has recently been explored for this purpose \cite{Hill2022}, \cite{Realisticdata2020,dorado-rojas_modelicagriddata_2023}. Such synthetic data is generated using extensive simulation physics-based power system models, which are challenging to maintain validated \cite{podlaski_validation_2022}. In this context, data augmentation arises as an attractive technique, to enlarge a pool of the relevant data for the DNN training, which has proven successful in computer vision \cite{nie2021resampling}, \cite{iglesias2023data}.

Ultra-fast performance of DNNs when deployed needs to be balanced against the training effort for the NN to perform well in real-world applications. Therefore, to speed up the process of learning and recycle the trained networks, transfer learning has been proposed. This concept suggests to recycle the trained DNNs when the data has a similar distribution, even though the NNs might be employed for slightly different tasks. This approach has been successfully applied in the image recognition and classification fields, for example, in classification of medical data, while in power system analysis it has been proposed to address event identification and dynamic security assessment tasks \cite{TL_Event2022}, \cite{TL_DSA2020}.

For power system small signal stability assessment (SSSA) the oscillatory pattern within the data can be used to determine the damping ratios of different modes \cite{osipov2022cross} and, therefore, the operational state of the system. To this end, data preprocessing methods \cite{luigi_preproc} are applied by grid operators' tools to detect unwanted behavior. 

In this work we propose to benefit from such processing step to enlarge the measurements of interest, use it to train the DNN and to direct the machine learning model to the features in the data that are important to learn. In addition we suggest to use the trained models by efficiently recycling them, i.e. use their parameters and their fixed structure for the repeated training process when new data arrive. Moreover, we evaluate if the preprocessing of the new collected data has a significant impact on the DNNs output. To this end, the variational mode decomposition (VMD) technique, that decouples the signal into meaningful modes, is exploited to augment data that is preprocessed via detrending.% By analogy to rotation, cropping, and flipping an image, the decomposed components of the signal may serve as a way to augment the limited data set with valuable dynamics. 
The verification methodology that is applied to check if the augmentation is successful is ``Train on Synthetic, Test on Real Data'' described in~~\cite{esteban2017real}.

The reminder of this paper is organized as follows. Section \ref{Problem} formulates a hypothesis. Section \ref{Data} presents the data augmentation process, the DNN to be trained and the DNN performance assessment metrics. Section \ref{DestrTest} describes the statistical tests for the hypothesis testing. Section \ref{CaseStudy} presents the evaluation of the proposed hypothesis on the SSSA task for power system data, such as voltage phasor angle. Finally, Section \ref{Conc} concludes this work.

%This document is a template for \LaTeXe. If you are reading a paper or
%PDF version of this document, please download the electronic file
%\texttt{ifacconf.tex}. You will also need the class file
%\texttt{ifacconf.cls}. Both files are available on the IFAC web site.

%Please stick to the format defined by the \texttt{ifacconf} class, and
%do not change the margins or the general layout of the paper. It
%is especially important that you do not put any running header/footer
%or page number in the submitted paper.\footnote{
%This is the default for the provided class file.}
%Use \emph{italics} for emphasis; do not underline.

%Page limits may vary from conference to conference. Please observe the 
%page limits of the event for which your paper is intended.

\section{Problem formulation}\label{Problem}

Variational Mode Decomponsition (VMD) is a method that extracts modes from a time-series data that has recently been explored in the analysis of power system modes \cite{osipov2022cross}. VMD allows to decompose the signal and extract a meaningful pattern in the form of Intrinsic Mode Functions (IMFs). The combination of these functions limits the bandwidth of the original signal, however, filters out the unnecessary information or noise. This ability of VMD may be utilized when training a DNN as a feature extraction step, thus shaping the pattern of the input training data to form another pull of data. Therefore, we wish to validate (or reject) the hypothesis if such data can be used as additional augmented data for training for the tasks where oscillations heavily influence the output of classification such as power system security assessment.

\begin{figure}
               \centering
                \includegraphics[width=0.48\textwidth]{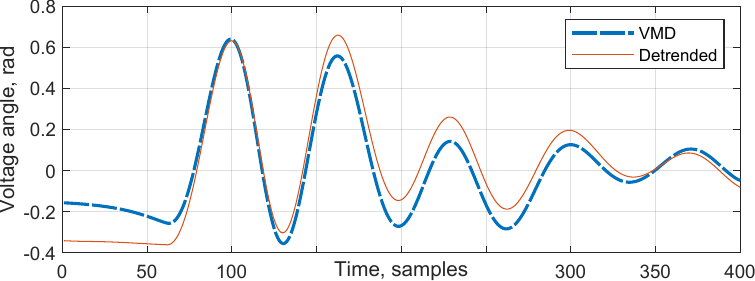}
                \caption{Examples of input signals}%: a) detrended (red); b) VMD decomposed (blue)}
                \label{DataSample}
\end{figure}

\textit{Hypothesis:} Can the VMD-decomposed data (e.g. as illustrated in Fig. \ref{DataSample}) be used as augmented data when the detrended data are used as input for a deep learning model, if the two distributions of the data are similar enough?

In this case if the hypothesis is validated, the augmented data can be considered as trusted augmentation since the method of data processing is known and unambiguous, which helps with data credibility.% contrary to recently emerging Generative Adversarial Networks outputs. 

\section{Data Generation and DNN Training}\label{Data}

The data to be used to train DNN is processed using detrending or VMD decomposition. When applying VMD, all components are summed up except the last component that contains nonstationary behavior. Detrending is a typical preprocessing step \cite{luigi_preproc}, however, the VMD is used here as an alternative since it has been shown to be effective in oscillation detection applications \cite{osipov2022cross}.

\subsection{Variational Mode Decomposition}
In this method the signal is decomposed into band-limited IMFs that are defined as:
\begin{equation}
    u_k(t) = A_k(t)cos(\phi_k(t))
    \label{IMF}
\end{equation}
where the phase $\phi_k(t)$ is a non-decreasing function ($\phi'_k(t) \geq 0$), the envelope is non-negative $A_k(t) \geq 0$, and the instantaneous frequency $\omega_k(t)$ varies much slower than the phase $\phi_k(t)$ \cite{dragomiretskiy2013variational}.

The constrained variational problem is the squared $L^2$-norm of the gradient: $\partial_t(\cdot)$:
\begin{multline}
    \min_{\{u_k\},\{\omega_k\}} \left(\sum_k \left\lVert \partial_t \left[\left(\delta(t)+\frac{j}{\pi t}\right)*u_k(t)\right]e^{-j\omega_k t} \right\rVert_2^2 \right)\\
    s.t. \sum_k u_k = f
    \label{Cost}
\end{multline}
where $\{u_k\} := \{u_1, ...,u_K\}$ and $\{\omega_k\}:= \{\omega, ..., \omega_K\}$ are the modes and their center frequencies, $\delta(t)$ is the Dirac distribution, $*$ denotes convolution.

The solution to \eqref{Cost} is obtained by utilizing the augmented Lagrangian multiplier with a series of iterative sub-optimizations known as the alternate direction method of multipliers (see \cite{dragomiretskiy2013variational}).

\subsection{Data Processing using Detrending} \label{sec:DataProc}
Voltage angle measurements naturally contain richer observability of modes that are excited after a contingency than other measurements \cite{irep}. Therefore, voltage angle measurements are used as input data for a deep learning model after pre-processing \cite{luigi_preproc}.

The following data preprocessing steps are performed for the voltage angle signal collected at the location of each electric power generator bus: a) subtraction of the center of angle that is defined as the inertia weighted average of all rotor angles \cite{tavora1972}; b) unwrapping; c) subtracting the initial value to obtain a deviation signal; d) linear detrending. 
Let the (raw) input data be defined as $\textbf{x}=[\textbf{x}_1, \textbf{x}_2, ..., \textbf{x}_T]$, where $T$ is the number of time-series signals. For the voltage angle data, the deviations of the signals are given by:
\begin{equation}
    \textbf{x}_{t} = [\angle{\theta_{1,t}}-\angle{\theta_{1,0}}, ..., \angle{\theta_{i,t}}-\angle{\theta_{i,0}}, \angle{\theta_{N,t}}-\angle{\theta_{N,0}}]
\label{eq:data_angle}
\end{equation}
where $N$ is the total number of buses, $\angle{\theta_{i,t}}$ is the voltage angle at bus $i$ of length $t$, and $\angle{\theta_{i,0}}$ is  the initial value of voltage angle for each bus $i$. 
%In addition, to test the ability of the deep learning algorithms to learn from noisy data, $1\%$ Gaussian noise has been added to the simulated signals that are used as pseudo-measurements in this work.

%In the case of the voltage angle, the data is presented as a set of vectors of $\textbf{x} = [\angle{\theta_{1,t}}, \angle{\theta_{2,t}},...,\angle{\theta_{N,t}}]$. 
Then, the angle unwrapping is performed by computing:
\begin{equation}    \angle{\theta_{j,i}}=\angle{\theta_{j,i}}+(2\pi k)\text{ if }(\angle{\theta_{j,i}}-\angle{\theta_{j,i-1}})\geq \pi
    \label{eq:angle_unwrap}
\end{equation}
where $j$ is the sample number in the data set, $i$ is the identifier of a measurement at a particular moment in time, and $k$ is a coefficient that is updated after every large jump in the phase value \cite{venkatasubramanian2016real}. Measurements are assumed to be collected at key system locations where Phasor Measurement Units are installed or synthetic data are obtained by simulating a power system model (as in this paper) \cite{irep}.
% Key system locations of interest are generator terminal buses at major power plants, major transmission level substations and boundary buses of tie-lines between the study and neighbouring systems.

In the last step of this process, i.e. labeling, VMD is applied and the IMF of the largest energy that contains the dominant modes is extracted. Then, Prony's method \cite{PronyJoe} is employed to the extracted IMF in order to identify modes. Next, calculate the damping of the mode closest to the critical system mode (i.e. $0.8\ Hz$ is the frequency of the inter-area mode in the examples herein).

In sum, the data generation and labeling approach using simulation models used herein is:
\begin{enumerate}
    \item Calculate the initial condition of the power system (i.e. obtain a power flow solution).
    \item Sample the contingency to be applied using realistic contingency generation \cite{Realisticdata2020} and simulate the behavior of the power system.
    \item The measured voltage angles in the buses of interest are pre-processed. The signal is detrended as shown in Fig. \ref{DataSample}.
    \item Label the trajectories identifying the state of the system. An example of the VMD decomposition and the detrended signal is shown in Fig. \ref{DataSample}
\end{enumerate}
\subsection{Offline Training} \label{sec:Offline}
To validate the formulated hypothesis (see Section \ref{Problem}), an Encoder \cite{serra2018towards}, a hybrid DNN architecture  that is designed for time-series data classification tasks, is chosen.
Encoder (Fig. \ref{Encoder}) consists of fully connected layers with an attention layer. Each of three convolutional layers includes respectively 128, 256, 512 filters, with the length of a 1D convolution window of sizes 5, 11, and 21 correspondingly. The operations within the hidden layers are given by:
\begin{figure*}
               \centering
                \includegraphics[width=0.80\textwidth]{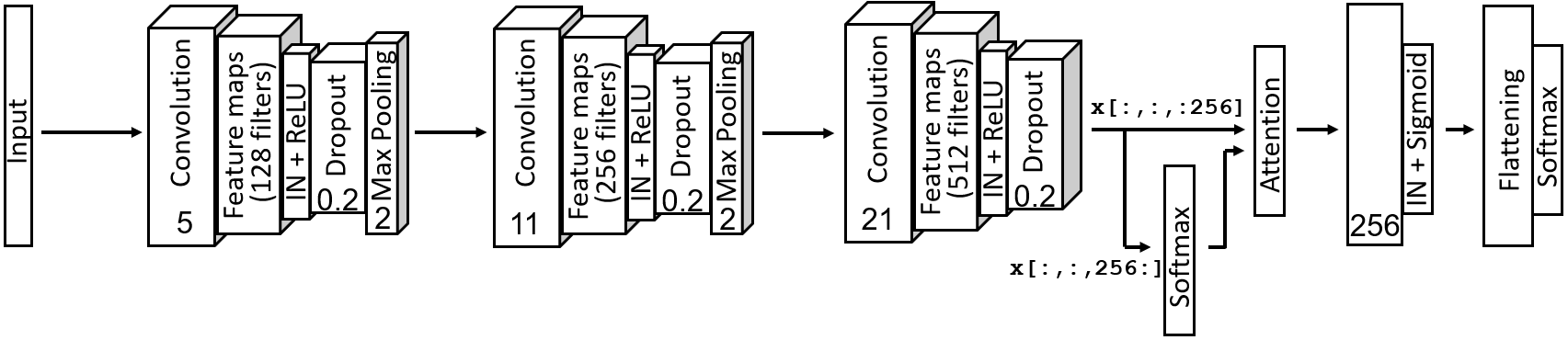}
                \vspace{-0.1in}
                \caption{Encoder neural network architecture}
                \label{Encoder}
\end{figure*}
\begin{equation}
\begin{aligned}
  x=Conv_{k_3}(Conv_{k_2}(Conv_{k_1}(x_0)))\\
  u = ATN(x[:,:,:256] \ast S(x[:,:,256:]))\\
  y = S(IN(\textbf{W} \otimes \textbf{u} + \textbf{b}))\\
 \end{aligned}
 \label{eq:Encoder}
\end{equation}
\noindent where $IN$ is the Instance Normalization operation, $k_1, k_2, k_3$ is the number of output filters; $ATN$ is the attention mechanism, $S$ is the softmax function. $IN$ normalizes and scales outputs of the previous layer. In contrast to batch normalization, this normalization does not operate on batches, but rather normalizes the activation of a single sample, making it suitable for recurrent neural networks \cite{ulyanov2016instance}. Meanwhile $ATN$ directs the NN to pay more attention to the small, significant parts of the data.
%consists of Keras lambda layer %\cite{Keraslambda} that is used to apply any function to the previous layer output. 
To implement $ATN$, the data is divided into two equal parts: $x[:,:,:256]$ and $x[:,:,256:]$ (see \eqref{eq:Encoder}). The softmax function is then applied to one of the parcels, and the two parcels are multiplied. This allows each element of the softmax-treated parcel to act as a weight for the other one. This mechanism enables the model to learn which parts of the time series are essential for the classification task.

The goal of the deep learning model is to minimize a loss function known as categorical cross entropy. The training set is composed of a collection of values $\{x_{(n)}, y_{(n)}\}_{n=1}^N$. The objective is to find the parameters of the model (e.g. \textbf{W}, \textbf{b} in equations (\ref{eq:Encoder})) that minimize the categorical cross entropy error function $L_{CE}$, and is given by
 \begin{equation}
 L_{CE}=\mathrm{min} \; \sum_{n=1}^{N} \sum_{c=1}^{C} y_{c,(n)} log(\hat{y}_{c,(n)})
 \label{CrossEntropy}
 \end{equation}
where $\hat{y}_{(n)}$ is the classification result of the input values ${x}_{(n)}$ for the trained model, $C$ is the number of classes, and $N$ is the number of training cases.

To solve \eqref{CrossEntropy}, the Adam optimizer was chosen with a learning rate equal to $0.00001$.
\subsection{Online Assessment}\label{sec:online}
For online SSSA, the previously trained Encoder is employed to classify the test cases using voltage angle measurements. With a set of bus voltage angle phasor measurements, it is straightforward to calculate the corresponding layer output values $h_1, h_2, ..., y_T$ given by \eqref{Encoder} using the trained NN parameters. To transform the output of the Encoder, $y_T \in (0, 1)$, into a stability index, a class separation threshold $\delta = 0.5 $ is defined so that the test cases with $y_T < \delta$ are considered stable and otherwise unstable.
\subsection{Evaluation Metrics}\label{sec:metrics}
%When the small signal assessment in defined as binary classification task, the input time-series sample is assigned to one of the classes. In this context the labels that are assigned to unstable and stable class are 1 and 0. %(positive and negative since one is interested to find the unstable scenarios).
%\textit{\textbf{Confusion Matrix}}
The training of the DNN has been performed using the accuracy metric, while precision and recall were measured on the testing data to validate the resulting model's quality. The class in which unstable cases belong is assigned as a positive class. The choice of positive class greatly influences on which class the precision and recall metrics will be focused on. 

\textit{\textbf{Accuracy}.} This metric defines a general performance of the model across all classes. %Usually this metric is most frequently used to evaluate a machine learning algorithm, however, it may become less relevant if the classes are not of equal importance.
\begin{equation}
    accuracy=\frac{N_{TP}+N_{TN}}{N_{TP}+N_{FP}+N_{TN}+N_{FN}}
\end{equation}
where $N_{TP}$ is the total of unstable cases correctly classified as unstable; $N_{TN}$ is the total of stable cases correctly classified; $N_{FP}$ is the number of stable cases misclassified as unstable; $N_{FN}$ is the number of unstable cases misclassified as stable.

\textit{\textbf{Precision.}} This metric evaluates accuracy of the model in classifying the data as a positive sample, which is, by our choice, an unstable case. Precision is defined as the ratio between true unstable samples and the total number of samples that are classified by the model as unstable, including those that are false unstable.

\textit{\textbf{Recall.}}
This metric evaluates the number of correct unstable predictions on all relevant unstable predictions. 
% \begin{equation}
%     recall=\frac{TP}{TP+FN}
% \end{equation}
\begin{equation}
    precision=\frac{N_{TP}}{N_{TP}+N_{FP}}, \ recall=\frac{N_{TP}}{N_{TP}+N_{FN}}
\end{equation}

In other words, recall evaluates missed correct predictions from the class that is labeled as unstable that is more important to be classified correctly. 

\textit{\textbf{Train on Synthetic, Test on Real Data (TSTR) \& Train on Real, Test on Synthetic Data (TRTS) \cite{esteban2017real}.}}
These are the techniques used to assess whether the augmented data (referred to as "synthetic data" in the names of the methods) are suitable for use as additional training data for the selected NN.
%Originally this method was proposed in  for validation of artificially generated realistic data using generative adversarial networks. 
For the purposes in this work, the approach can be applied in two ways: when the VMD-decomposed data is used for training but the original detrended data is applied for testing, and vice versa. The resulting evaluation metrics, such as accuracy, recall, and precision, are compared with those received in the original data.

\section{Distributions similarity Kernel Maximum Mean Discrepancy test}\label{DestrTest}
To verify whether the VMD-decomposed data are good to serve as augmented data for NN training, the input data distribution similarity test has to be applied. The Kernel Maximum Mean Discrepancy (KMMD) is one of the recent statistical tests developed to determine if two samples of input data are drawn from different distributions and has been effective in machine learning applications \cite{gretton2006kernel}.

The maximum mean discrepancy (MMD) is a measure of similarity between two distributions with given observations $X = \{x_1, ..., x_m\}$ and $Y = \{y_1, ..., y_n\}$. It is determined by using a function from the class $\mathcal{F}$. Let $\mathcal{F}$ be a unit ball in a universal reproducing kernel Hilbert space $\mathcal{H}$, defined on the compact metric space $\mathcal{X}$ with associated kernel $k(\cdot, \cdot)$, then $MMD[\mathcal{F},X,Y]$ is defined as:
\begin{equation}
\begin{aligned}
    MMD[\mathcal{F},X,Y]= \Big[ \frac{1}{m^2} \sum_{i,j=1}^{m} k(x_i, x_j) - \frac{2}{mn} \sum_{i,j=1}^{m, n} k(x_i, y_j) +\\
    \frac{1}{n^2} \sum_{i,j=1}^{n} k(y_i, y_j) \Big] ^{\frac{1}{2}}
\end{aligned}
\label{eq:MMDemp}
\end{equation}
where $k(x,x') = \langle \phi(x), \phi(x') \rangle$ - kernel in kernel Hilbert space. The most common kernel function that is applied in this test is the Radial Basis kernel function (Gaussian). The similarity metric $MMD[\mathcal{F},X,Y] = 0$ if two distributions are equal $X=Y$.

The uniform convergence bound for the empirical MMD (see \eqref{eq:MMDemp}) that defines a threshold of the hypothesis test is based on the Rademacher complexity \cite{cortes2013learning} and presented below under the assumptions that $m=n$, $|k(x,y)|\leq K$, distributions are the same:
\begin{equation}
\begin{aligned}
    MMD_b[\mathcal{F},X,Y]> m^{-\frac{1}{2}}\sqrt{2\textbf{E} [k(x, x)-k(x,x')]}+\epsilon \\
    >2(K/m)^{1/2}+\epsilon
\end{aligned}
\end{equation}
both with probability less than $exp(-\frac{\epsilon^2 m}{4K})$. Thus, a hypothesis test of level $\alpha$ for the null hypothesis $X=Y$ (i.e. $MMD[\mathcal{F},X,Y] = 0$) has acceptance region $MMD_b[\mathcal{F},X,Y] < \sqrt{2(K/m)}(1+\sqrt{2log\alpha^{-1}})$ according to Corollary 16 in \cite{gretton2008kernel}. The boundary of this region is later named Rademacher. This boundary is relaxed compared to the asymptotic boundary of the unbiased estimate of $MMD_u^2$. According to Corollary 18 in \cite{gretton2008kernel}, a hypothesis test of level $\alpha$ for the null hypothesis that two distributions are the same $X=Y$ has the acceptance region $MMD_u^2 < (4K/\sqrt{m})\sqrt{log(\alpha^{-1})}$. The boundary of this region is later named the asymptotic boundary.

\section{Case Studies and Analysis} \label{CaseStudy}
Several case studies have been developed to test the hypothesis of using VMD-decomposed data as augmented data for SSSA. The KMMD distribution similarity test is applied in two variants of the relaxed Rademacher boundary and the tight asymptotic boundary. The test is validated on the same distribution data that were split into two sets, on the original detrended data and randomly generated, and on the original and the augmented set to test the hypothesis.
If the VMD-decompoded data come from a distribution similar to the original detrended voltage angle phasor data, additional DNN performance metrics are computed for the original and the augmented data. These are TSTR and TRTS metrics that are explained in Section \ref{sec:metrics}.

\subsection{Testing distributions similarity using Kernel Maximum Mean Discrepancy Test}
\begin{figure*}
                \centering
                 \includegraphics[width=0.85\textwidth]{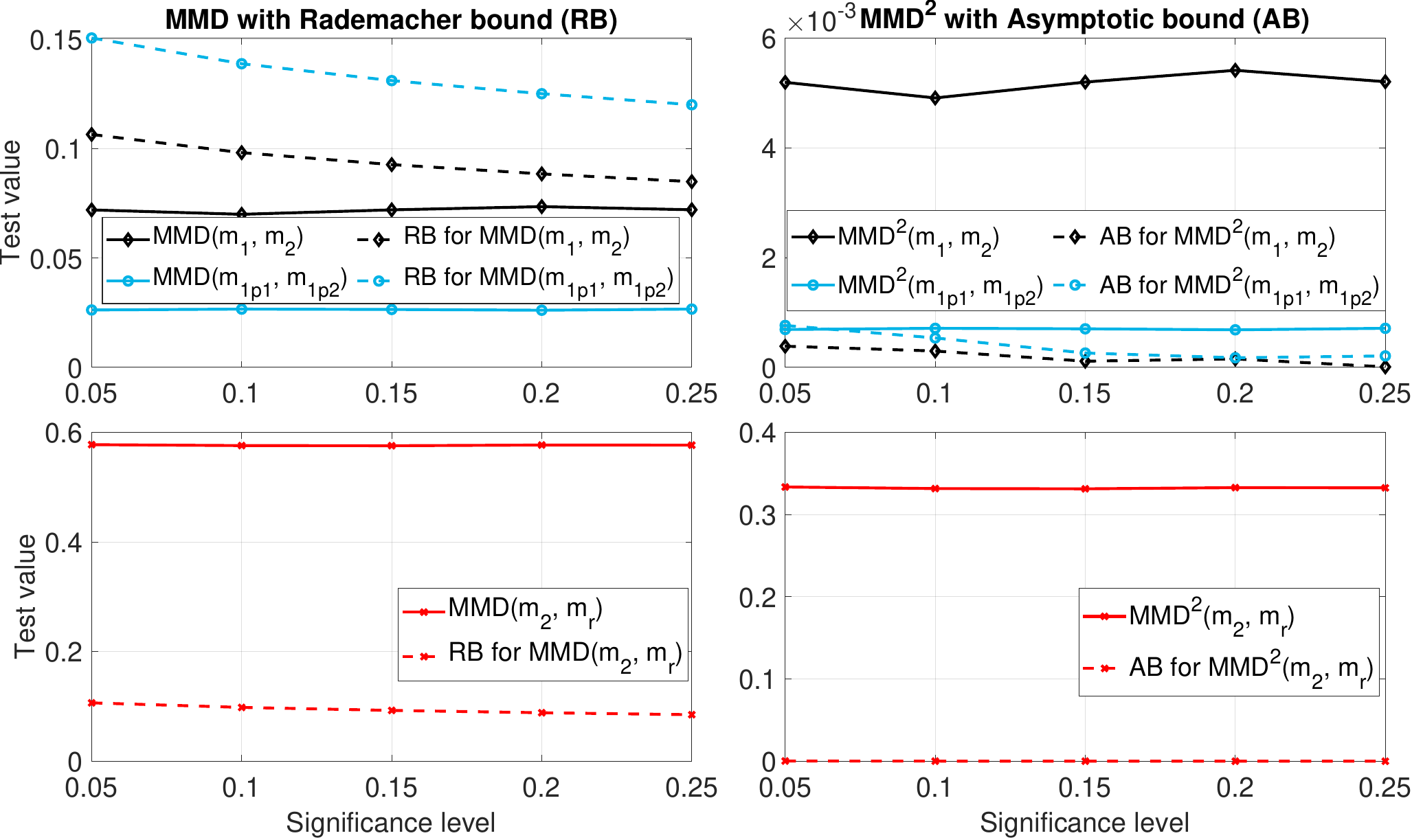}
                 \caption{Testing distributions similarity using Kernel Maximum Mean Discrepancy Test with the Rademacher and Asymptotic bound. $m_1$ - the decomposed data set, $m_2$ - the detrended voltage angle phasor data set,  $m_r$ - randomly generated data set, $m_{1p1}, m_{1p2}$ - two parts of the same data set $m_1$ %, MMD refers to the metric of the similarity KMMD test (see equation (\ref{eq:MMDemp}))
                 }
                 \label{fig:DataComp}
 \end{figure*}

Distribution similarity testing results comparing the detrended and decomposed voltage angle data are presented in Fig. \ref{fig:DataComp} and Table \ref{tab:True_False}. Each data set consists of 7878 samples with each sample length of 400 points with the recording rate $60$ samples per second. 

The KMMD similarity tests are performed as detailed in Section \ref{DestrTest}. The MMDs calculated with the bounds are presented in the same color showing the correspondence with the same test result. When the MMD value is smaller than the boundary value, the hypothesis rejection is considered true, otherwise false. The tests are evaluated for different confidence levels that correspond to the significance level $\alpha$. The meaning of significance level is the probability to reject hypothesis when it is true. In Figure \ref{fig:DataComp} the case of comparison of the distributions of the decomposed data and the randomly generated data is used as a baseline. In this case, the values of $MMD$ or $MMD^2$ are significantly higher than the relaxed Rademacher and tight asymptotic boundaries, respectively. Thus, the hypothesis that the data originate from the same distribution is rejected. 

Another test is performed to validate the proposed methodology on the detrended data that is divided into two parts. Both tests with the relaxed and tight bound have shown that the hypothesis cannot be rejected for all levels of significance. Thus, both case studies reflect the expected performance of the distribution similarity tests.

Finally, the test result on the detrended and VMD-decomposed data using the Rademacher boundary is that the hypothesis cannot be rejected. In other words, the distributions are locally similar enough to consider the data to be of close origin. The VMD-decomposed data can be used as augmented data. However, the test with the asymptotic boundary resulted in the hypothesis being rejected. Therefore, considering the different test results (see Table \ref{tab:True_False}), deep learning model performance validation is performed using the TSTR and TRTS metrics.

\begin{table}
\large
\caption{The hypothesis testing results}
\resizebox{\columnwidth}{!}{
\centering
\begin{tabular}%{P{0.01\textwidth} P{0.04\textwidth} P{0.02\textwidth} P{0.09\textwidth} P{0.09\textwidth} P{0.09\textwidth}}
{c| c| c| c}
\hline
 Data Set 1 & Data Set 2 &  \multicolumn{1}{|p{3cm}|}{\centering $MMD_b$ \\ with Rademacker bound} &  \multicolumn{1}{|p{3cm}}{\centering $MMD_u$ \\ with Asymptotic bound} \\%[0.4ex]
\hline
Detrended & Detrended &  Not Rejected & Not Rejected \\ \hline
 Decomposed & Randomly generated & Rejected & Rejected  \\ \hline
 Detrended & Decomposed & Not Rejected & Rejected  \\
\hline
% \multicolumn{5}{p{0.5\textwidth}} {where $\angle{\theta}$ - voltage angle, $V$ - voltage magnitude. For all the cases the data are generated using \cite{Realisticdata2020}, the data length is 10 sec.}
\end{tabular}
}
\label{tab:True_False}
\end{table}

\subsection{Training of the deep learning model Encoder on detrended and decomposed data of the 769-bus power system model} \label{VA769Data}
%To conclude on possibility to employ the VMD-decomposed voltage angle data as the augmented data, 
To address the main hypothesis in this work (see Section \ref{Problem}), the metrics TSTR and TRTS (see Section \ref{sec:metrics}) are employed. In addition, accuracy, precision and recall are employed to evaluate the Encoder's performance on different training and testing data. The idea is that if the data sets both contain the main features that distinguish the state of the system, and the difference in the distributions is not significant, the resulting performance do not change or change in an acceptable range. The case study has been carried out on 5252 training samples and tested on 2626 samples. The results are summarized in Table \ref{tab:TSTR_TRTS_Metrics}. The Encoder's training results for the solely detrended or decomposed data are very similar in terms of accuracy and recall, holding the largest difference of 1 \% in precision value. The intuition behind this difference is that the ratio between the number of false unstable cases and true unstable cases is larger when the Encoder is trained on the decomposed data. However, for the other two cases that use different data for training and testing, the change in performance of the Encoder is more prominent, especially in precision and recall metrics values. %Encoder mistaken stable cases for the unstable case when trained on the decomposed data.
Although the difference between cases with the same data and different data for training and testing in accuracy is around 1 \%, the precision differs up to 5 \% and the recall is up to 7 \%. The biggest drop in performance is for the case where the training is performed on the VMD decomposed data, meaning less rich data than the original detrended data set. This result is logical if the Encoder learns fewer patterns during the training than it is present in the test data.

\begin{figure}[ht!]
               \centering
                \includegraphics[width=0.38\textwidth]{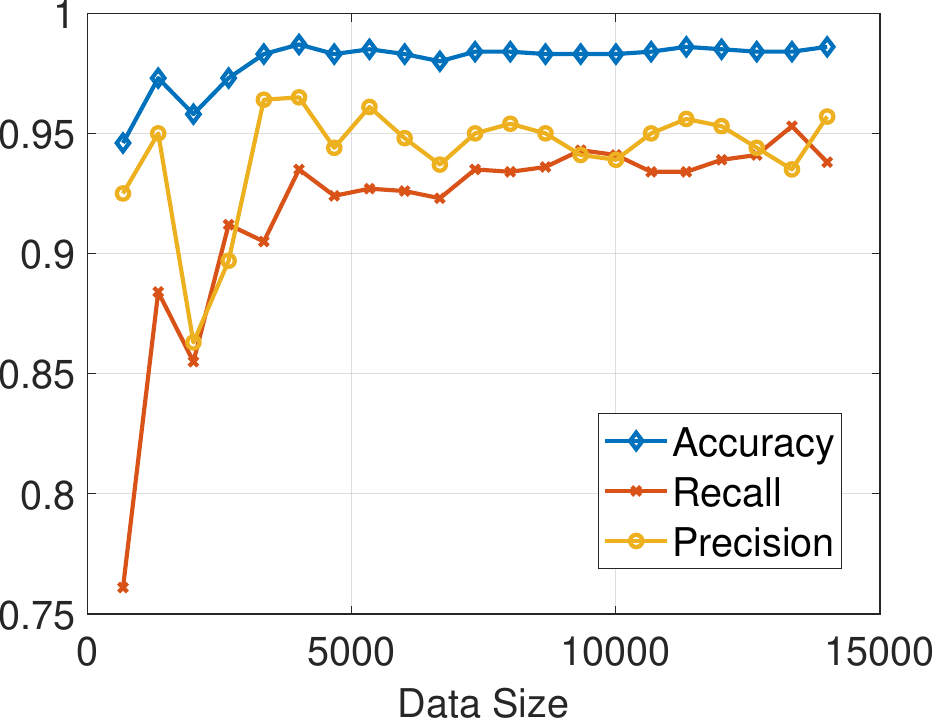}
                \caption{Performance of the Encoder on data size}
                \label{fig:DataSize}
\end{figure}

\begin{table}
\large
\caption{Performance of the Encoder using TSTR and TRTS metrics}
\resizebox{\columnwidth}{!}{
\centering
\begin{tabular}%{P{0.01\textwidth} P{0.04\textwidth} P{0.02\textwidth} P{0.09\textwidth} P{0.09\textwidth} P{0.09\textwidth}}
{c| c| c| c| c}
\hline
 & \multicolumn{1}{p{3cm}|}{\centering Train: Detrended, \\ Test: Detrended} & \multicolumn{1}{|p{3cm}|}{\centering Train: Decomposed, \\ Test: Decomposed} & 
 \multicolumn{1}{|p{3cm}|}{\centering Train: Detrended, \\ Test: Decomposed} &
 \multicolumn{1}{|p{3cm}}{\centering Train: Decomposed, \\ Test: Detrended}\\%[0.4ex]
\hline
Accuracy, \% & 98.71 & 98.51 & 97.33 & 97.22 \\ \hline
Precision, \% & 97.07 & 96.03 & 91.53 & 92.80 \\ \hline
Recall, \% & 93.62 & 93.30 & 89.25 & 86.70 \\
\hline
% \multicolumn{5}{p{0.5\textwidth}} {where $\angle{\theta}$ - voltage angle, $V$ - voltage magnitude. For all the cases the data are generated using \cite{Realisticdata2020}, the data length is 10 sec.}
\end{tabular}
}
\label{tab:TSTR_TRTS_Metrics}
\end{table}
Finally, a study on the merged original and augmented data is carried out to observe the performance of the Encoder depending on the size of the data set. In Figure \ref{fig:DataSize} the performance on the joint data (the original detrended and the augmented VMD-decomposed data) has shown a stable high performance when the data size is larger than 5500 samples. Therefore, even though the data sets are not completely from identical distributions according to the statistical test results in Table \ref{tab:True_False}, the VMD-decomposed data merged with the original data set of the voltage angle data gives as good performance as the base case with the detrended data.

% \section{Discussion} \label{Discussion}

% Time series data augmentation is performed in different ways - bending and shifting a signal in a way the signal processing allowed preserving the main characteristics of the data that are meaningful for the particular task. 
% %allows to decide where to draw the boundary between classes to clearly and correctly relate the input data with the appropriate category in the classification problem
% Thus, pursuing the main goal of these transformations, the developed methodology is aimed to preserve the validity of the decision making when  employing the deep learning model. Thus, ensuring the data is sampled from the similar distribution and at the same time the data can be trusted by the engineers is a two-folded problem that is addressed. The results shown that the distribution of the VMD-decomposed data is similar to the original detrended voltage angle data under relaxed Rademacher boundary. Even though the tight asymptotic boundary did not allowed the hypothesis of the distributions similarity to be accepted, the validation of the deep learning neural network performance on the augmented data using the Train on Synthetic, Test on Real Data and Train on Real, Test on Synthetic Data metrics shown a good performance when training the Encoder. This allows to conclude that the VMD-decomposed data can be used as the augmented data. Additionally, the VMD decomposition is a trusted method of data processing which allows to filter out the unnecessary frequencies narrowing a bandwidth of the signal. 

\section{Conclusion} \label{Conc}

In this paper we propose to use variational mode decomposition to produce the augmented data for training the deep neural network to perform a small signal assessment for a large power system. By analogy to rotation, cropping, and flipping of an image, the decomposed components of the signal may serve as a means to augment the limited data set with valuable dynamics.

To validate the hypothesis that VMD-decomposed data can serve as augmented data for the neural network, the KMMD statistical test of distribution similarity is performed. The results of the testing have shown that the VMD-decomposed data can be considered as augmented data under relaxed conditions posed by the Rademacher boundary. The additional validation of the proposed augmentation has been performed using TSTR and TRTS metrics. 

The outcome has shown good performance of the Encoder neural network used on the newly formed data. The Encoder has demonstrated a consistently positive performance with the increasing data size, which includes the additional ``augmented'' data.

\bibliography{ifacconf}             % bib file to produce the bibliography
                                                     % with bibtex (preferred)

%\appendix
%\section{A summary of Latin grammar}    % Each appendix must have a short title.
%\section{Some Latin vocabulary}              % Sections and subsections are supported  
                                                                         % in the appendices.
\end{document}